\documentclass{article}

\usepackage{graphicx}
\usepackage{amsmath}
\usepackage{cite}

\begin{document}

\title{LAMMPS Implementation of Rapid Artificial Neural Network Derived Interatomic Potentials}

\author[1]{D.Dickel$^{1,2}$, M. Nitol$^{2}$, C. D. Barrett$^{1,2}$}

\begin{abstract}
While machine learning approaches have been successfully used to represent interatomic potentials, their speed has typically lagged behind conventional formalisms. This is often due to the complexity of the structural fingerprints used to describe the local atomic environment and the large cutoff radii and neighbor lists used in the calculation of these fingerprints. Even recent machine learned methods are at least 10 times slower than traditional formalisms. An implementation of a rapid artificial neural network (RANN) style potential in the LAMMPS molecular dynamics package is presented here which utilizes angular screening to reduce computational complexity without reducing accuracy. For the smallest neural network architectures, this formalism rivals the modified embedded atom method (MEAM) for speed and accuracy, while the networks approximately one third as fast as MEAM were capable of reproducing the training database with chemical accuracy. The numerical accuracy of the LAMMPS implementation is assessed by verifying conservation of energy and agreement between calculated forces and pressures and the observed derivatives of the energy as well as by assessing the stability of the potential in dynamic simulation. The potential style is tested using a force field for magnesium and the computational efficiency for a variety of architectures is compared to a traditional potential models as well as alternative ANN formalisms. The predictive accuracy is found to rival that of slower methods.

\end{abstract}

\maketitle

\section{Introduction}
\label{1_Intro}

Artificial neural networks (ANNs) and other machine learning techniques are gaining prominence as a method for developing classical interatomic potentials for use in molecular statics (MS) and molecular dynamics (MD) simulation \cite{zong2018developing, ArtrithUrbanTiOANN2016,sosso2012neural,behler2014representing,dragoni2018achieving,seko2015first,huan2017universal,deringer2017machine,bartok2018machine,kobayashi2017neural}. Because of their ability to accurately fit large and complex data sets and because they can be rapidly optimized, ANNs have shown the potential to predict atomic energies and forces more accurately than most existing classical formalisms \cite{seko2015first,huan2017universal}. By using density functional theory (DFT), which is relatively expensive computationally, to produce large training databases for the ANN, potentials which reproduce this data to high precision at much faster computational speeds can be created. While only a limited number of large scale dynamic simulations have been performed using these potentials, static simulations for a variety of materials have been able to reproduce \textit{ab initio} forces and energies with chemical accuracy \cite{TakahashiMLIPTi2017,ArtrithUrbanTiOANN2016,PunSiGAP2017,dragoni2018achieving}.\\
ANNs used to model interatomic potentials behave similarly to traditional formalisms with an energy being defined for each individual atom based on the local atomic arrangement and the atomic species of it and its neighbors, mapping the local environment directly to numerical values for the energy and forces.  This encoding of the local environment is referred to as the structural fingerprint. Thompson \textit{et al.} developed the Spectral Neighbor Analysis Method (SNAP) \cite{thompson2015spectral} which uses machine-learning techniques on components of the local neighbor density. Behler and Parrinello \cite{BehlerANN2011,behler2007generalized} defined a set of functions based on Gaussians which could be used to generate the fingerprint. This fingerprint formalism has since been used by a number of authors as the basis for an ANN. The celebrated Gaussian Approximation Potentials (GAP)\cite{bartok2010gaussian} have been used to develop potentials with chemical accuracy for a variety of systems including tungsten \cite{szlachta2014accuracy}, lithium-carbon\cite{fujikake2018gaussian}, iron\cite{DragoniFeGAP2018} and water\cite{nguyen2018comparison}. Artrith and Urban used the Behler-Parrinello basis functions to create a potential for titanium dioxide using an ANN with two hidden layers with 10 neurons each which accurately described the various cystal phases of TiO\textsubscript{2}\cite{ArtrithUrbanTiOANN2016}. Artrith, Urban, and Ceder have also recently proposed an efficient scheme for the generation of machine-learned potentials for composite materials with many atomic species\cite{artrith2017efficient}. Takahahsi \textit{et al} \cite{TakahashiMLIPTi2017} used a combination of the Gaussian formalism along with insights from the modified embedded atom method (MEAM) \cite{Baskes92} to generate a fingerprint of several thousand values which was then used to fit a potential for titanium using linear ridge regression. Recently, a method using the embedding functions from MEAM has been used to generate potentials for titanium and zinc using a single hidden layer \cite{dickel2020neural,nitol2021artificial}. Singraber, Behler, and Dellago \cite{singraber2019library} recently presented the \textit{n2p2} neural network package which uses the Behler symmetry functions in the construction of the fingerprint. This package has been used to develop a number of potentials including for Al-Cu \cite{marchand2020machine} and Mg \cite{stricker2020machine}.\\
While the optimal structural fingerprint and network architecture for an ANN is still unclear and indeed is probably strongly dependent on the material system being examined, there is a need to be able to perform large scale dynamic calculations with these potentials. Additionally, while machine learned potentials greatly outperform \textit{ab initio} calculations in terms of efficiency, they still lag behind conventional formalisms such as the embedded atom method (EAM)\cite{DawBaskes84} and MEAM, which have been the standard for analysis of metals since their introduction over 25 years ago. Additionally, MEAM takes advantage of angular screening to limit the size of the neighbor list and incorporate the known shielding effect to improve performance. To this end, we have created an implementation of a rapid artificial neural network (RANN) potential formalism for use in LAMMPS \cite{Plimpton95LAMMPS}. This implementation accepts arbitrary Multi-layer Perceptron (MLP) architectures and structural fingerprints of arbitrary size. Sample potentials have been generated from the MEAM based structural fingerprints with angular screening, though novel functions or fingerprint types can be added independently. Section 2 of this paper reviews the formalism for ANN interatomic potentials and provides the specific structural fingerprints implemented including a discussion of the use of angular screening. Section 3 describes the framework which has been implemented in LAMMPS. Section 4 demonstrates a number of simple static and dynamic calculations using an ANN potential. In particular, conservation of energy is shown along with accurate calculation of the forces and pressure, and a comparison is done between the computational efficiency of this potential style as compared to other common styles for metal systems. It is demonstrated that, for the network architectures capable of reproducing a magnesium training database to within 1meV/atom, performance approximately one third as MEAM can be achieved. Additionally, the performance of the potential as a function of neural network architecture and neighbor list size is considered.

\section{Artificial Neural Network Architecture}
\label{2_primer}

\begin{figure}%
\center
\fbox{\includegraphics[scale=0.5]{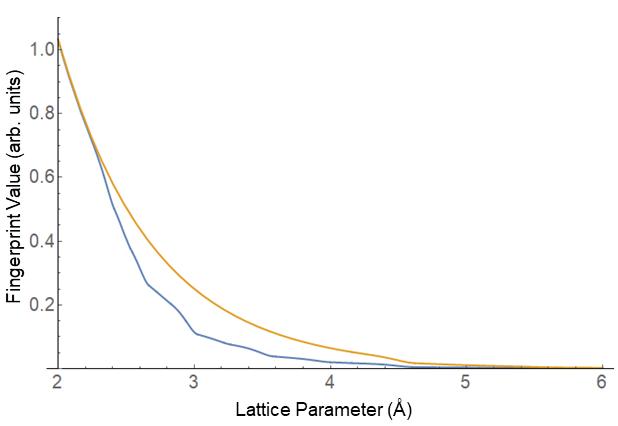}}%
\caption{A sample value for a fingerprint in the presence (orange) and absence (blue)of angular screening as the volume of a cell is changed isotropically. Because the neighbor list frequently changes without angular screening, the curve is seen to be less smooth, requiring more data to provide predictive information to the ANN.}
\label{fig:screen}%
\end{figure}

Most machine-learned interatomic potentials make use of MLPs to predict the energy of individual atoms in a system given their environment. Details of the training and optimization of these networks in reference to \textit{ab initio} data can be found elsewhere \cite{ArtrithUrbanTiOANN2016,dickel2020neural,behler2016perspective}, but the general structure of a network as an interatomic potential is as follows. The energy of a particular atom $i$, determined by its environment, is the last of N layers of the neural network. The values for any particular layer, $\mathbf{A}^n$, after the first is determined by the previous layer and the weight and bias matrices $\mathbf{W}^n$ and $\mathbf{B}^n$:
\begin{eqnarray}
Z^n_{l_n}=\sum_{l_{n-1}}{W^n_{l_nl_{n-1}}A^{n-1}_{l_{n-1}}+B^n_{l_n}} \\
A^n_{l_n}=g^n(Z^n_{l_n})
\end{eqnarray}
Where $l_n$ is the number of neurons in layer $n$ and $g^n(x)$ is a nonlinear activation function. The input layer, $\mathbf{A}^0$ is given by the structural fingerprint of the local atomic environment. The ouput layer, $\mathbf{Z}^N$, will always contain a single node, representing the energy, and so $\mathbf{W}^N$ will always be a row vector and $\mathbf{B}^N$ always a single number.\\
A number of different functions can be used to describe this structural fingerprint. Behler and Parrinello \cite{behler2007generalized} consider two types of fingerprint functions: radial symmetry functions which are a pairwise sum of Gaussians, and angular terms over all triplets of atoms. For the RANN style, however, we use the fingerprint style introduced by Dickel, Francis, and Barrett, motivated by the Modified Embedded Atom Method (MEAM) formalism \cite{Baskes92,Baskes97,Baskes2NN2000}, with the addition of angular screening. This style seems particularly relevant for metal potentials for the same physically motivated reasons that the MEAM formalism is effective. In this style, two different kind of input fingerprints are considered. First, simple pair interactions are considered and summed over all the neighbors of a given atom. For a given atom labeled $i$, we define a set of pair potentials interactions with the form
\begin{equation}
F_n=\sum_{j\neq i}{(\frac{r_{ij}}{r_e})^ne^{-\alpha_n\frac{r_{ij}}{r_e}}f_c(\frac{r_c-r_{ij}}{\Delta r})S_{ij}}
\end{equation}
Where $j$ labels all the neighbors atoms of $i$ within a cutoff radius $r_c$, $n$ is an integer, different for each member of the pairwise contributions to the fingerprint, $r_e$ is the equilibrium nearest neighbor distance, $S_{ij}$ is an angular screening term described below, and $\alpha_n$ are metaparameters, which can be tuned to better optimize the potential. In principle, $\alpha_n$ are related to the dimensionless $\alpha$ used in the MEAM formalism which is related to the bulk modulus of the material \cite{Baskes2NN2000}. $f_c(x)$ is a cutoff function which smoothly varies the weight of the interaction from 1 for atoms inside the cutoff radius to 0 once $r_{ij}>r_c$ where $\Delta r$ defines the width of the transition. The cutoff function used here is the same as employed in MEAM and is given by
\begin{equation}
r_c(x)= \left\{\begin{array}{lr}
1, & x> 1\\
(1-(1-x)^4)^2, & 0\leq x \leq 1\\
0, & x< 0
\end{array}\right\}
\end{equation}
The second kind of fingerprint function considers three body terms, with a form similar to the partial electron densities used in MEAM. There are two parameters, $m$ and $k$ which can be varied to generate more inputs for the fingerprint:
\begin{equation}
G_{m,k}=\sum_{j,k}{\cos^m{\theta_{jik}}e^{-\beta_k\frac{r_{ij}+r_{ik}}{r_e}}f_c(\frac{r_c-r_{ij}}{\Delta r})f_c(\frac{r_c-r_{ik}}{\Delta r})S_{ij}S_{ik}}
\label{eq:threebody}
\end{equation}
where $\theta_{jik}$ is the angle between $r_{ij}$ and $r_{ik}$, m is a non-negative interger and $\beta_k$ is a set of metaparameters which determine the length scale of the different terms. For large numbers of neighbors, it can be more efficient to calculate this as a sum over a single list of neighbors as follows.
\begin{eqnarray}
G_{0,k}=\left(\sum_j{e^{-\beta_k\frac{r_{ij}}{r_e}}f_c(r_{ij}})S_{ij}\right)^2\\
G_{1,k}=\sum_{\alpha_1}\left(\sum_j\frac{r^{\alpha_1}_{i,j}}{r_{i,j}}e^{-\beta_k\frac{r_{ij}}{r_e}}f_c(\frac{r_c-r_{ij}}{\Delta r})S_{ij}\right)^2\\
G_{2,k}=\sum_{\alpha_1,\alpha_2}\left(\sum_j\frac{r^{\alpha_1}_{i,j} r^{\alpha_2}_{i,j}}{r_{i,j}^2}e^{-\beta_k\frac{r_{ij}}{r_e}}f_c(\frac{r_c-r_{ij}}{\Delta r})S_{ij}\right)^2\\
G_{3,k}=\sum_{\alpha_1,\alpha_2,\alpha_3}\left(\sum_j\frac{r^{\alpha_1}_{i,j} r^{\alpha_2}_{i,j} r^{\alpha_3}_{i,j}}{r_{i,j}^3}e^{-\beta_k\frac{r_{ij}}{r_e}}f_c(\frac{r_c-r_{ij}}{\Delta r})S_{ij}\right)^2\\
G_{m,k}=\sum_{\alpha_1}\sum_{\alpha_2}...\sum_{\alpha_m}\left(\sum_j\frac{r^{\alpha_1}_{i,j} r^{\alpha_2}_{i,j}...r^{\alpha_m}_{i,j}}{r_{i,j}^m}e^{-\beta_k\frac{r_{ij}}{r_e}}f_c(\frac{r_c-r_{ij}}{\Delta r}))S_{ij}\right)^2
\label{eq:threebodyastwo}
\end{eqnarray}
where $r^{\alpha_m}_{ij}$ is the $x$, $y$, or $z$ component of $r_{ij}$ for $\alpha_m=1,2$ or $3$, respectively. A simplification can be made by noting that many of the terms in the three body interactions are redundant. In $G_{2,k}$ for example, there will be a calculation for xy as well as for yx. By taking advantage of this redundancy, we can reduce the number of terms that need to be calculated for $G_{m,k}$ from $3^n$ to $\frac{(n+1) (n+2)}{2}$. The simplified expression for $G_{m,k}$ now reads:
\begin{equation}
\label{eq:twobody}
G_{m,k}=\sum_{\alpha_1\geq\alpha_2\geq...\geq\alpha_m}\left(\frac{n!}{n_x!n_y!n_z!}\sum_j{\frac{r^{\alpha_1}_{i,j} r^{\alpha_2}_{i,j}...r^{\alpha_m}_{i,j}}{r_{i,j}^m}e^{-\beta_k\frac{r_{ij}}{r_e}}f_c(\frac{r_c-r_{ij}}{\Delta r})}\right)^2
\end{equation}
where $n_x$, $n_y$, and $n_z$ are the total number of x-, y-, and z-components in the set $(\alpha_1,\alpha_2,...,\alpha_m)$, respectively. Note that which of these forms (Eq. \ref{eq:threebody} or Eq.\ref{eq:twobody} will be more efficiently calculated will depend on the length of the neighbor list and the magnitude of $m$. Structural fingerprints of arbitrary size can be constructed by varying the number of different $n$, $m$, and $k$ used giving more or less information to the ANN. For the potentials tested here, we have used consecutive values for n and m and totaling $n_{tot}$ and $m_{tot}$ respectively ($n\in(-1,0,1,...,n_{tot}-2)$, $m\in(0,1,...,m_{tot}-1)$). A particular ANN potential will consist of a fixed structural fingerprint, number and length of each hidden layer, weight and bias matrices for each layer and activation functions for each layer. For the potentials considered here we have used the following activation functions:
\begin{eqnarray}
g^N(x)=x \label{eq:activation1}\\
g^n(x)=\frac{x}{10}+\frac{9}{10}\log(e^x+1)\; \textrm{for} \;n<N
\label{eq:activation2}
\end{eqnarray}
The size of the weight and bias matrices will depend on the length of the fingerprint and the length of each hidden layer.\\
As discussed above, one of the major computational barriers in the implementation of ANN methods is the large neighbor list included in the calculation of the structural fingerprint. In order to obtain good convergence with DFT results, cutoff radii used can be larger than 12 \AA \cite{stricker2020machine}, which for the ground state of Mg includes over 300 atoms. By comparison, recent MEAM potentials for Mg \cite{wu2015magnesium} have used a cutoff distance of 5.875 \AA, which includes only 38 atoms in the ground state. Since the computation time scales at least linearly with the number of neighbors, minimizing the length of the neighbor list without sacrificing performance should be a key goal for optimal efficiency. The radial screening function $f_c$ can accomplish this, but leads to nonphysical results at large separations (see, for example \cite{wu2015magnesium}). Angular screening, whereby the effective interaction between atoms is reduced or eliminated by the presence of an atom located between them can effectively limit the neighbor list in a similar way without the unphysical results. Such a method has been utilized by MEAM \cite{baskes1997determination}, and the same screening method has been employed here. Briefly, the screening between two atoms is determined from the product of all the screening interactions by other atoms in the neighborhood:
\begin{equation}
S_{ij}=\prod_{k\neq i,j}S_{ikj}
\end{equation}
where $S_{ikj}$ is calculated from a geometric construction considering the ellipse formed by the 3 atoms with $r_{i,j}$ one of the axes. The screening parameter, $C_{ikj}$ is then given by:
\begin{equation}
C_{ikj}=1+2\frac{r^2_{ij}r^2_{ik}+r^2_{ij}r^2_{jk}-r^4_{ij}}{r^4_{ij}-(r^2_{ik}-r^2_{jk})^2}
\end{equation}
and then screening value is
\begin{equation}
S_{ikj}=f_c\left(\frac{C_{ikj}-C_{min}}{C_{max}-C_{min}}\right)
\end{equation}
where $f_c$ is the same cutoff function used for the radial cutoff and $C_{max}$ and $C_{min}$ are metaparameters which can be tuned to determine which neighbors can be excluded from calculations.\\
The effect of including angular screening can be demonstrating by considering the change of an individual fingerprint as the length scale is changed continuously. For this example, we consider the value of a single fingerprint for a Mg atom in a perfect bulk hcp lattice as the lattice parameter is changed continuously. In the absence of angular screening, the value of the fingerprint will change rapidly at particular values of the lattice constant as new neighbors enter the radial screening distance. If angular screening is included with metaparameters such that, for example, only 3rd nearest neighbors are ever included regardless of lattice parameter, the change in the value is considerably smoother. Figure \ref{fig:screen} demonstrates the difference between the two cases. Not only is the unscreened value more expensive to calculate as it requires more neighbors for most cases, it can also be seen that the predictive power of the model is hampered as the relation between two states which are physically similar -- ideal 3D lattices with different lattice constants, given markedly different trends and values as the number of neighbors changes. This can be overcome through the action of the neural network and precise cancellation among different fingerprint terms, but the quality of fit and the extrapolative power of the screened functions appears more natural. Indeed unphysical deviations at relatively moderate isotropic compressions, as seen for example in \cite{stricker2020machine} can be avoided entirely through angular screening as the fingerprint values change continuously in the screened case even at high compression.\\
For the full network, forces are calculated by taking a gradient of the energy over the entire system. This includes the individual ANN contributions from every atom in the system, so the force on atom $i$ is determined not only by the ANN output of that atom, but of all of its neighbors as well.\\

\section{Numerical Implementation}
\begin{figure}%
\center
\fbox{\includegraphics[scale=0.6]{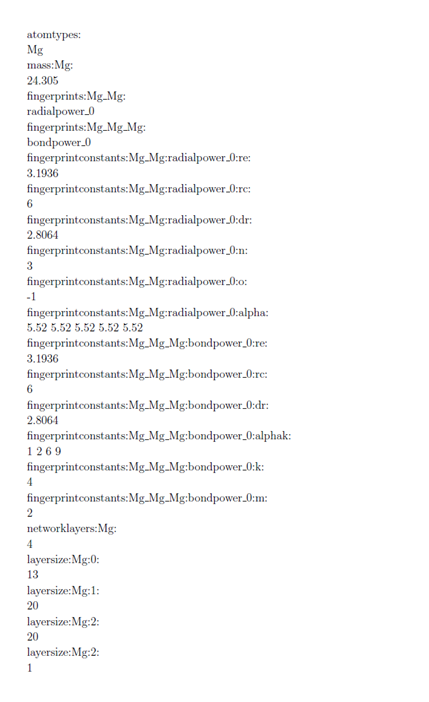}}%
\caption{Sample header for the neural network potential style. The header specifies the different element types as well as the style and number of fingerprints used for each type. Metaparameters needed for each style are also included as well as the overall network architecture.}
\label{fig:header}%
\end{figure}

The ANN potential file written for LAMMPS includes all of the details of the neural network. It begins with a header which lists the types and numbers of fingerprints to be used. While we have only included the two varieties, the bond power and radial power terms, $F_n$ and $G_{m,k}$, in this work, the LAMMPS potential style can be expanded in a straightforward way to include other descriptors. For each of the types used, the potential file then specifies the number used and any metaparameters. For $F_n$ this includes the range of values for $n$ as well as the metaparameters $\alpha_n$, $r_c$, $C_{max}$, $C_{min}$ and $r_e$. For $G_{m,k}$, $m_{tot}$ and $k_{tot}$ are given along with the metaparameters $\beta_k$, $r_e$, $r_c$, $C_{min}$, and $C_{max}$. Note that while we use the same cutoff radius and angular screening parameters for both varieties, they can in principle be different as every fingerprint type has its own set of metaparameters. Next, the network architecture is specified by giving the total number of layers and the length of each layer. Figure \ref{fig:header} shows a sample header for the potential file. Following the header is the weight and bias matrices for each layer and finally the activation functions used for each layer. Using the potential file, the ANN subroutine then calculates the feature list for every atom and uses these to calculate the energy and force for every atom in the usual way. The subroutine and potential file also have the capability of assigning multiple ANNs to different atoms types to be used for multi-element simulations. In this way, the framework can be used for arbitrary structural fingerprints as inputs to arbitrary MLP architectures for systems with an arbitrary number of element types.\\
The subroutine is included as a normal pair-style in LAMMPS called `rann' and can be called in the usual way:\\

\noindent pair\textunderscore style rann\\
pair\textunderscore coeff * * potential\textunderscore file.nn Element1 Element2 ...

\section{Validation and Benchmarking}
\label{4_test}

With the pair style implemented in LAMMPS, several potentials were fit for a reference database for magnesium (Mg). The database, containing over 300,000 atomic environments was constructed using the open-source density functional theory (DFT) software, Quantum Espresso \cite{giannozzi2009quantum}. The configurations used to train the networks are shown in Table \ref{tab:1}. These configurations included simple triaxial strain perturbations of the low energy SC, FCC, BCC, and HCP phases of Mg along with larger supercells with atoms displaced from equilibrium to mimic the effect of thermal excitation. Free surface data and vacancy data, with atoms also perturbed from equilibrium positions were also included to improve stability. Effective temperatures up to 1000K were included in the thermally excited database. Training of the network was carried out by minimizing the RMSE of the system energies compared to the DFT results using the Levenerg-Marquadt method \cite{Levenberg1944,Marquardt1963}. 10\% of the training data from each configuration type was left out of training and used for validation. The most significant variation however, came in the size and construction of the fingerprint used as the input layer. Along with the number and type of fingerprints included, the most important factor in determining accuracy and efficiency was the number of atoms included in the neighbor list. This is a function of the cutoff radius $r_c$ and the angular screening parameter $C_{min}$. Table \ref{tab:2} summarizes the different fingerprints considered and the RMSE for both the training and validation datasets for each potential. As can be seen, even the smallest architectures had RMSE of less than 3 meV/atom and the best were less than 1 meV/atom. Table \ref{tab:2} also includes computational performance for a sample calculation which will be described in detail below.

\begin{table}[htbp]
  \caption{DFT simulations used to generate training and validation data for the Al potential. Training database attempts to consider all relevant low energy structures which may be encountered during MD simulations.}
  \scalebox{0.7}{
    \begin{tabular}{cccc}
    Sample Description & Atoms per simulations & Number of Simulations & Total atomic environments\\
    \hline
    SC cubic cell w/ strains up to $\pm$15\% & 1 & 500 & 500\\
    FCC primitive cell w/ strains up to $\pm$15\% & 1 & 500 & 500\\
    BCC primitive cell w/ strains up to $\pm$15\% & 1 & 500 & 500\\
    HCP unit cell w/ strains up to $\pm$10\% & 2 & 2700 & 5400\\
    \hline
    FCC 2x2x2 orthogonal supercell & 32 & 500 & 16000\\
    BCC 3x3x3 orthogonal supercell & 54 & 500 & 27000\\
    HCP 3x3x3 primitive supercell & 54 & 2000 & 108000\\
    HCP 4x3x3 primitive supercell with vacancy & 67 & 100 & 6700\\
    HCP {0001} free surface & 54 & 500 & 27000\\
    HCP {$10\overline{1}0$} free surface & 54 & 100 & 5400\\
    Totals & & 8100 & 307000\\
    \end{tabular}}%
  \label{tab:1}%
\end{table}%

\begin{table}[htbp]
  \centering
  \caption{Input Fingerprints considered in this work. $k_{tot}$ and $m_{tot}$ are the total number of values uses for $k$ and $m$ in Eq. \ref{eq:twobody}. For the pair-style input, $n_{tot}$ was always 5. All ANNs contained a single hidden layer of 20 neurons. \#SF is the total number of structural fingerprints for a particular potential and $r_c$ and $C_min$ are the radial cutoff and angular screening parameter that determine the number of neighbors considered in calculation of the fingerprints. \#NN is the number of neighbors included in a calculation of bulk HCP Mg at the equilibrium lattice constant. RMSE values are for the full training and validation databases and are given in meV/atom. Calculation speed is given in timesteps/second. The starred entry was used for validation tests.}
    \begin{tabular}{ccccccccc}
    $k_{tot}$ & $m_{tot}$ & \#SF & $r_c$ & $C_{min}$ & \#NN & Training RMSE & Validation RMSE & Calculation speed\\
    *3 & 4 & 13 & 6 & 0.49 & 20 & 0.90 & 2.25 & 21.77\\
    3 & 4 & 13 & 8 & 0.49 & 20 & 0.67 & 1.64 & 15.74\\
    3 & 4 & 13 & 10 & 0.49 & 20 & 0.69 & 1.71 & 10.71\\
    2 & 4 & 13 & 12 & 0.25 & 38 & 0.89 & 1.71 & 5.56\\
    2 & 4 & 13 & 12 & 0.49 & 20 & 1.09 & 1.05 & 8.26\\
    2 & 4 & 13 & 12 & 0.70 & 18 & 0.93 & 1.08 & 9.34\\
    3 & 4 & 17 & 12 & 0.25 & 38 & 0.80 & 1.03 & 5.07\\
    3 & 4 & 17 & 12 & 0.49 & 20 & 0.71 & 1.38 & 8.04\\
    3 & 4 & 17 & 12 & 0.70 & 18 & 0.80 & 1.06 & 8.75\\
    5 & 3 & 20 & 12 & 0.25 & 38 & 1.57 & 2.95 & 4.91\\
    5 & 3 & 20 & 12 & 0.49 & 20 & 2.80 & 3.08 & 7.88\\
    5 & 3 & 20 & 12 & 0.70 & 18 & 2.88 & 3.10 & 8.32\\
    MEAM \cite{wu2015magnesium} & & & & & & & & 61.963 \\
    N2P2 \cite{stricker2020machine} & & & & & & & & 5.12 \\
    \end{tabular}%
  \label{tab:2}%
\end{table}%

\subsection{Validation of the potential}

Tests of the validity of the potential and its implementation  in LAMMPS were all performed using the 13 member input fingerprint with a single hidden layer containing 20 neurons. The metaparameters for this potential are shown in Table \ref{tab:meta}. $C_{min}$ was chosen exclude atoms beyond the third nearest neighbors in the equilibrium hcp Mg structure. First, to confirm agreement with first principles results beyond the RMSE in energy, the bulk properties of hcp magnesium were tested for this potential. Table \ref{tab:3} shows a comparison of these bulk properties among the sample ANN, the first principles database, and experimental results. Agreement can be seen on the simple bulk properties targeted.\\

\begin{table}[htbp]
  \centering
  \caption{Metaparameters of the benchmark potential used for validation.}
    \begin{tabular}{cc}
      $n_{tot}$ & 5 \\
      $k_{tot}$ & 3 \\
      $m_{tot}$ & 4 \\
      $\alpha$ & 5.52 \\
      $\beta_{k}$ & 1,2,9 \\
      $r_c$ & 6 $\AA$ \\
      $C_{min}$ & 0.49 \\
      $C_{max}$ & 2.90 \\
      \end{tabular}%
  \label{tab:meta}%
\end{table}%

\begin{figure}%
\center
\includegraphics[scale=0.5]{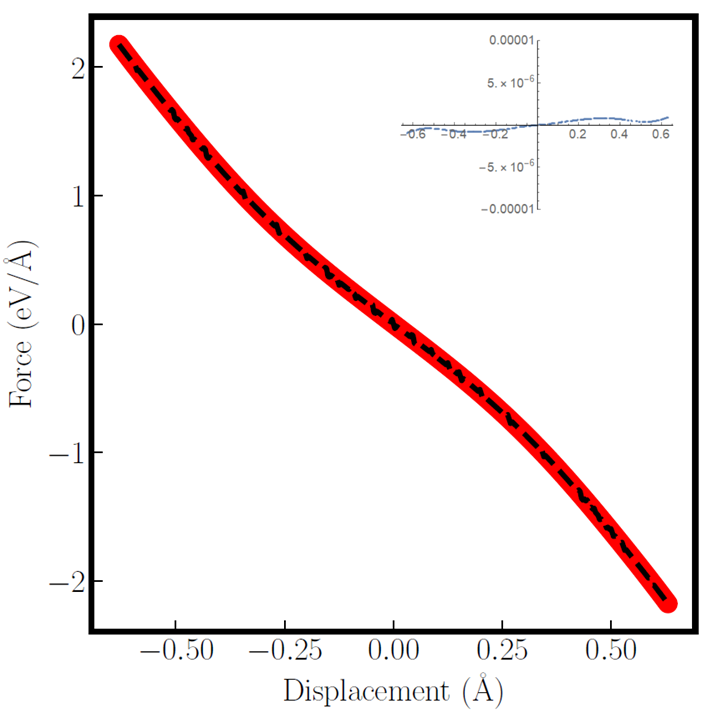}%
\caption{The force as determined by the numerical derivative of the system energy (solid red) and the calculated force (dashed black) as a single atom is displaced in the ($\overline{1}\overline{1}20$) direction in a 4x4x4 hcp Mg lattice. (inset) Difference between derivative and direct force calculation.}
\label{fig:1}%
\end{figure}

As an initial test of the fidelity of the implementation, a number of simple structures were generated and tested to confirm that the forces and energies were correct and in agreement with one another. As a simple test, we displace a single atom from its equilibrium location in a minimized HCP Mg lattice. Figure \ref{fig:1} shows the force on the displaced atom as a function of displacement as predicted by the ANN and by numerical differentiation of the energy. Agreement is seen between the two measures demonstrating that forces are correctly predicted by the algorithm.\\

\begin{table}[htbp]
  \centering
  \caption{Properties of bulk magnesium as determined by experiment, the DFT database used for the fitting of the potential, and the 13-20-1 neural network used in this study.}
    \begin{tabular}{cccc}
    Property & Experiment & DFT & ANN \\
    $a$ (nm) & 3.209 \cite{saunders1988metastable} & 3.194 & 3.196 \\
    $c$ (nm) & 5.211 \cite{saunders1988metastable} & 5.184 & 5.189 \\
    $c/a$ (nm) & 1.623 & 1.623 & 1.623 \\
    $E_{c}$ (eV) & 1.51 \cite{saunders1988metastable} & 1.45 & 1.46 \\
    $C_{11}$ (GPa) & 63.5 \cite{simmons1971single} & 61.6 & 63.4 \\
    $C_{12}$ (GPa) & 25.9 \cite{simmons1971single} & 23.8 & 24.3 \\
    $C_{13}$ (GPa) & 21.7 \cite{simmons1971single} & 21.2 & 18.5 \\
    $C_{33}$ (GPa) & 66.5 \cite{simmons1971single} & 64.3 & 62.9 \\
    $C_{44}$ (GPa) & 18.4 \cite{simmons1971single} & 17.3 & 18.4 \\
    \end{tabular}%
  \label{tab:3}%
\end{table}%

\begin{figure}%
\center
\includegraphics[scale=0.5]{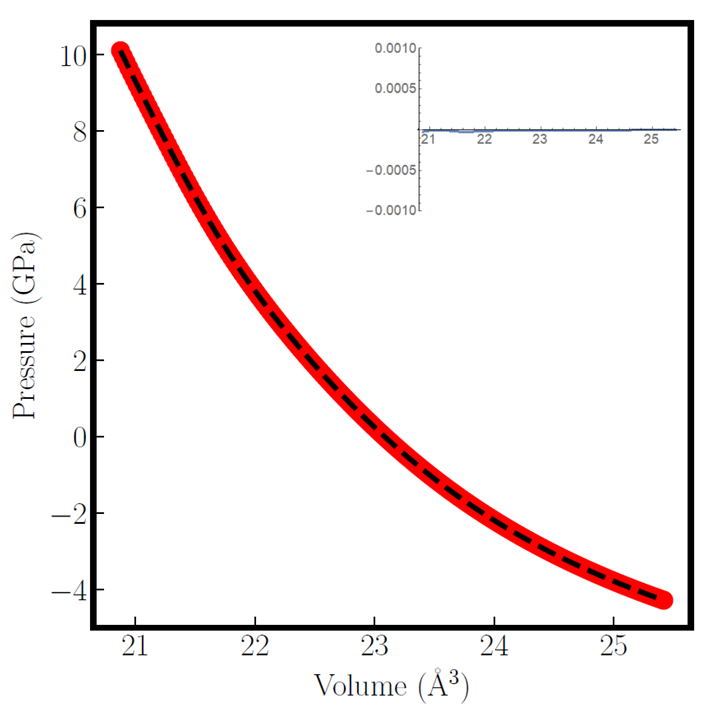}%
\caption{The xx component of the stress tensor (solid red) and the derivative of the system energy (dashed black) as uniaxial tension/compression is applied in the x direction. (inset) Difference between derivative and stress calculations.}
\label{fig:2}%
\end{figure}

\begin{figure}%
\center
\includegraphics[scale=0.5]{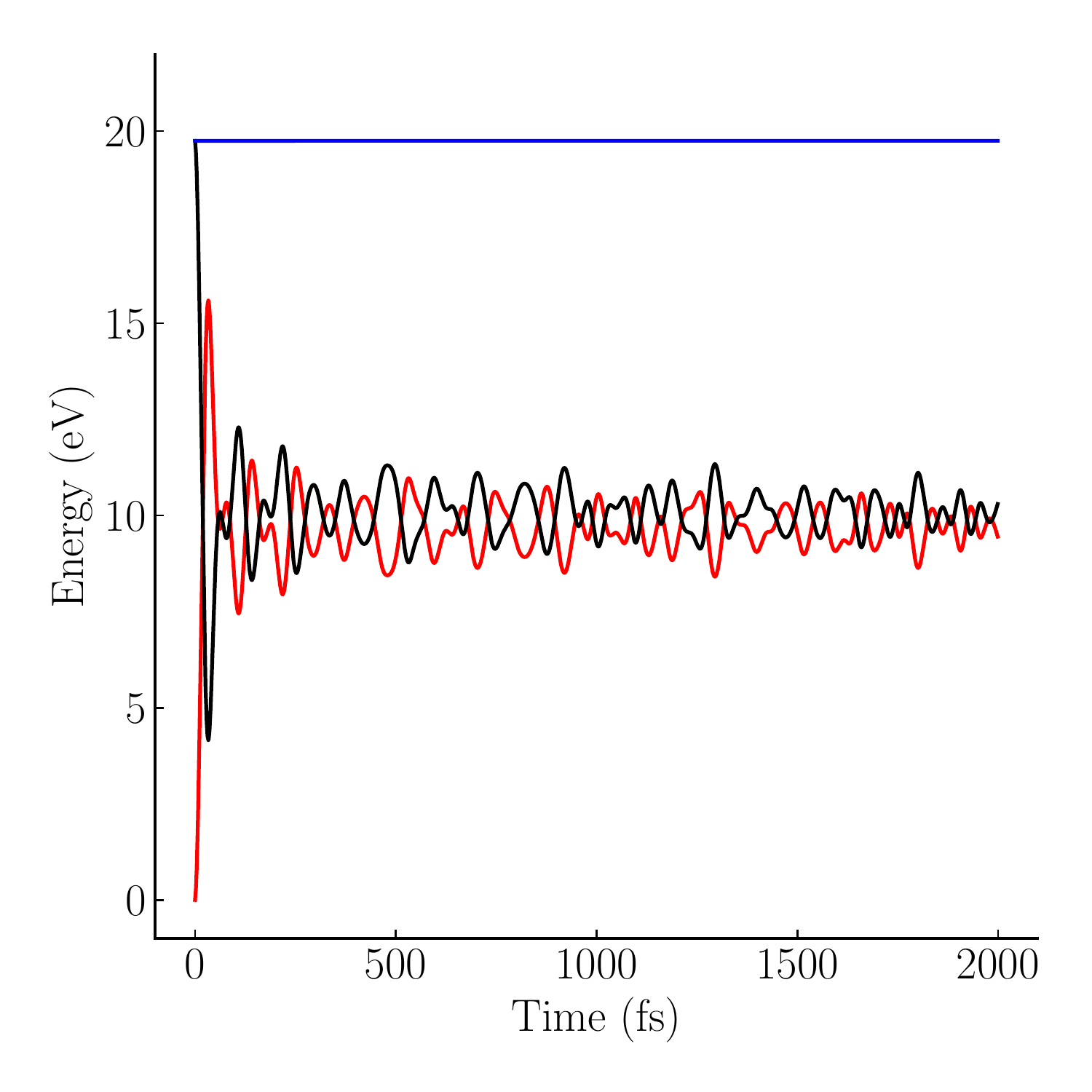}%
\caption{The potential, (red) kinetic, (black) and total (blue) energy relative to the 0K minimum for a 4x4x4 hcp lattice of magnesium atoms with initial velocity distribution producing a temperature of 300K. The system was allowed to evolve under NVE conditions. Conservation of energy can be observed for this system.}
\label{fig:3}%
\end{figure}

As a second test, a periodic fcc cubic unit cell was generated and allowed to relax following a conjugate gradient minimization scheme. Symmetry is maintained by the structures throughout minimization. The simulated cell is then uniaxially compressed 10\% in the x direction and slowly pulled in tension to a total positive strain of 10\% and the changes in energy and stress tensor are examined. The results are shown in Figure \ref{fig:2}. We see, again, that the numerical derivative of the energy as a function of strain agrees exactly with the stress calculated by the algorithm.\\
Having demonstrated that the pair style correctly reproduces the energy, forces, and stresses in static configurations, we now confirm that dynamic behavior can also be correctly produced. A dynamic simulation was carried out with a 256 atom periodic hcp system (4x4x4 orthogonal unit cells). The system was initialized from the equilibrium ground state with atoms given a random velocity distribution producing an average temperature of 300K. The system was then allowed to evolve at constant energy and volume. Figure \ref{fig:3} shows the evolution of the kinetic, potential and total energy of the system. We see that total energy is conserved.

\subsection{Computational Efficiency}

While ANN potentials have the capability to reproduce experimental or first principles results with greater accuracy than traditional formalisms, if the computational performance suffers too greatly, the benefits become negligible. The speed at which the ANN pair style will perform depends on the size and complexity of the structural fingerprint, the number of neighbors within the cutoff radius and the network architecture.\\
As a benchmark of the performance, we compare the performance of the Mg ANN potential to an magnesium Modified Embedded Atom potential\cite{wu2015magnesium} and a recent n2p2 style neural network potential \cite{stricker2020machine}. To test the speed of the MEAM potential and the various NN potentials, a sample system of hcp magnesium containing 4000 atoms was minimized and each atom given a random velocity matching a distribution with a temperature of 300K. The systems was then allowed to evolve at constant energy and volume for 1000 timesteps. The number of timesteps calculated per second is shown in Table \ref{tab:2}. Calculations were performed on one core of a Intel Xeon E5-2690 2.0GHz processor. We see that all of the potentials are slower than MEAM, with a strong dependence on the number of neighbors considered and on the cutoff radius $r_c$, but the fastest potential, used in the validation study, runs approximately 1/3 as fast as MEAM and four times faster than competitive n2p2 potentials. As can be seen the first columns of Table \ref{tab:2}, there is a strong dependence on the cutoff radius even when the same number of atoms are included due to the angular screening parameter. This is due to the computational time required to calculate the angular screening needed to remove atoms from the neighbor list. This is still seen to be more efficient than considering all the atoms within the larger cutoff radius.\\
For the size of networks considered here, the computation time is dominated by the calculation of the structural fingerprint with the propagation through the network only a limiting factor for the smallest fingerprints and largest architectures. This suggests that, assuming the structural fingerprint contains sufficient information to describe the relevant configurations, it is more efficient to increase the size of the network to increase the accuracy of the potential, using this formalism, rather than expanding the fingerprint. This is particularly true for extending the length of the cutoff radius. While a large cutoff radius is often desirable and is used in existing ANN potentials as it should improve the smoothness of the potential as atoms move away from the target atom, the high computational cost of increasing the neighbor list suggests that ideally only first, second, and possibly third neighbors should normally be considered for metals. Angular screening combines both of these aspects, limiting the number of neighbors in bulk settings while still allowing relatively long-range interactions for studies involving voids, free surfaces, or fracture.\\
The RANN formalism presented here has the advantage of requiring similar computer time as MEAM to run simulations using the developed potentials and, thanks to the additional of angular screening, faster than other existing ANN formalisms, but the development time should be considerably smaller, as fitting to an arbitrary number of targets is automated with many free parameters available to tune to all available data.\\



\section{Conclusions}
\label{5_Conc}

We have presented an implementation of the Rapid Artificial Neural Network (RANN) interatomic potential pair style for use in the LAMMPS molecular dynamics code. The potential is suitable for static and dynamic calculations, conserving total energy and correctly calculating the pressure and forces of bulk solids. Such potential styles have been shown to be capable of reproducing training data from first principles calculations at a level which exceeds previous semi-empirical formalisms \cite{seko2015first, huan2017universal,kobayashi2017neural}. The developed pair style is capable of accommodating network architectures of arbitrary dimension and activation function, and the calculation of the structural fingerprint requires only singly looped summations over the neighbor list of a target atom, improving computational efficiency. Angular screening has also been introduced both to improve efficiency by limiting the number of included neighbors and to improve predictive power by introducing the physically motivated phenomenon of shielding and smoothing fingerprints as atoms move through the radial cutoff. Computational efficiency is of particular importance for this formalism as the improvements in accuracy over existing formalisms such as MEAM are greatly diminished if the runtime is closer to that of first principles calculations. As such, we demonstrate that for a network with 13 input fingerprints and a cutoff radius of 6.0 $\AA$ for the neighbor list, and angular screening restricting interactions to third nearest neighbors in bulk HCP the implementation performs at one third the speed of MEAM. Larger networks, fingerprints, or neighbor lists will hamper this performance, with the most significant cost associated with increasing the neighbor list through the length of the radial cutoff. Ultimately, a flexible, scalable formalism for ANNs, which can utilize the native parallel processing available in LAMMPS has been demonstrated with a formalism which produces high accuracy potentials which operate on speeds comparable to those of MEAM.\\

\section*{Acknowledgements}

\section*{References}

\bibliographystyle{unsrt}
\bibliography{Bibliography}{}

\end{document}